\documentclass{emulateapj}
\usepackage{apjfonts}

\shorttitle{ALMA GRB Host Galaxies}
\shortauthors{Wang, Chen, \& Huang}

\begin{document}

\title{ALMA Submillimeter Continuum Imaging of the Host Galaxies of GRB\,021004 and GRB\,080607}

\author{Wei-Hao Wang\altaffilmark{1}, Hsiao-Wen Chen\altaffilmark{2}, and Kui-Yun Huang\altaffilmark{1}}

\altaffiltext{1}{Academia Sinica Institute of Astronomy and Astrophysics, P.O. Box 23-141, Taipei 10617, Taiwan}
\altaffiltext{2}{Department of Astronomy \& Astrophysics and Kavli Institute for Cosmological Physics, University of Chicago, Chicago, IL 60637, USA}

\begin{abstract}
We report 345 GHz continuum observations of the host galaxies of 
gamma-ray bursts (GRBs) 021004 and 080607 at $z>2$ using the Atacama Large Millimeter/Submillimeter
Array (ALMA) in Cycle 0.  Of the two bursts, GRB\,021004 is one of the few GRBs that originates in a
Lyman limit host, 
while GRB\,080607 is classified as a ``dark burst'' and its host galaxy is a candidate of dusty star 
forming galaxy at $z\sim3$.  
With an order of magnitude improvement in the sensitivities of the new imaging searches, we detect the 
host galaxy of GRB\,080607 with a flux of $S_{345} = 0.31\pm 0.09$ mJy and a corresponding 
infrared luminosity of $L_{\rm IR}=(2.4$--$4.5)\times10^{11}~L_\sun$.   However, the host galaxy of GRB\,021004 
remains undetected and the ALMA observations allow us to place a 3-$\sigma$ upper limit of 
$L_{\rm IR}<3.1\times10^{11}~L_\sun$ for the host galaxy.
The continuum imaging observations show that the two galaxies are not
ultraluminous infrared galaxies but are at the faintest end of the dusty galaxy population that gives rise to 
the submillimeter extragalactic background light. The derived star formation rates of the two GRB host galaxies 
are less than 100 $M_\sun$~yr$^{-1}$, which are broadly consistent with optical 
measurements.  The result suggests that the large extinction ($A_V\sim3$) in the afterglow of 
GRB\,080607 is confined along its particularly dusty sightline, and not representative of the
global properties of the host galaxy.
\end{abstract}

\section{Introduction}
Long-duration gamma-ray bursts (GRBs) are believed to originate in the death of massive stars 
(see \citealp{woosley06} for a recent review), and are thus expected to trace star formation in
galaxies \citep[e.g.,][]{wijers98,totani99}.  Because of the extreme luminosity of the prompt
emission and the afterglow, GRBs are a powerful probe of star 
formation in early times \citep[e.g.,][]{tanvir09,salvaterra09}.  In order to establish the link between 
GRBs and the cosmic star formation, it is important to understand the properties of the GRB host
galaxies \citep{hjorth12}. A critical measurement is the star formation rates (SFRs)
of the host galaxies. 

There exist various SFR indicators for high-redshift galaxies in different spectral windows.
One key concern is the effect of dust extinction. Even with arguably good
extinction corrections in optical data, highly obscured components may still exist and 
would only appear at the far-infrared and radio wavelengths. 
The presence of such components would indicate a significant spatial variation in dust content 
in which case a global extinction correction would not apply.
For GRB host galaxies, systematic surveys were carried out to observe continuum emission in the
radio \citep{michalowski12} and submillimeter \citep{berger03,tanvir04,priddey06} frequency ranges for constraining 
dust enshrouded SFR. Because of the synchrotron spectral slope,
radio observations are only effective in detecting GRB host galaxies at $z\lesssim1$ \citep[e.g.,][]{michalowski12}.
Dust continuum emission in the submillimeter has a spectral slope that can nearly cancel the effect of 
luminosity distance from $z\sim1$ to $z\sim10$, making the submillimeter wavelengths an effective window for 
detecting faint galaxies at high redshifts \citep{blain93}.
However, the 850 $\mu$m survey of 21 GRB host galaxies at $z<3.5$ by \citet{tanvir04} 
using the James Clerk Maxwell Telescope (JCMT) only uncovered three host galaxies at $>3\sigma$ 
confidence levels, and all three hosts are at $z<1.5$.  
Submillimeter single-dish telescopes are confusion limited at roughly a few 
mJy at 850 $\mu$m, and therefore can only detect galaxies with infrared luminosities of 
$L_{\rm IR}$ (8 to 1000 $\mu$m)~$> 10^{12.5}~L_\sun$, or SFR $\gtrsim1000~M_\sun$~yr$^{-1}$. 
This SFR  limit is  much larger than the typical SFR of GRB host galaxies measured in
the optical \citep{christensen04,savaglio09}. Deeper submillimeter measurements 
are thus required to better
constrain their infrared luminosity and the underlying SFR.

A particularly interesting class of GRB is ``dark GRBs'' \citep{djorgovsky01,jakobsson05}, defined by
their faint optical afterglow, relative to the bright X-ray emission. A  definition
for dark bursts is those with the optical-to-X-ray spectral index of $\beta_{\rm OX} < 0.5$ \citep{jakobsson04},
which is physically motivated based on theoretical predictions of the synchrotron model.
Approximately 30\%--50\% of long-duration GRBs have suppressed optical fluxes relative to their X-ray emission
\citep{melandri08,cenko09,melandri12}. The weaker optical emission can be caused by either intergalactic medium
absorption at $z>6$ 
\citep{kawai06,greiner09,tanvir09,salvaterra09} or dust extinction in their host galaxies \citep{perley09}.  
In the latter case, dark GRBs may serve as a tracer of dust enshrouded star formation across cosmic time. 
However, uncertainty remains regarding whether the observed dust obscuration is representative of the 
global properties of the host galaxies or merely local to the progenitor site.  This uncertainty can be 
addressed by comparing the rest-frame infrared luminosities between dark GRB host galaxies and the 
rest of the host galaxy population.

In this \emph{letter}, we present initial results from a pilot study of GRB host galaxies in the submillimeter
frequency range using the Atacama Large Millimeter/Submillimeter Array (ALMA). In Cycle 0, we 
observed the host galaxies of GRB\,021004 ($z=2.330$) and GRB\,080607 ($z=3.036$) at 345 GHz.
The GRB fields were selected to have early-time afterglow spectra available for constraining the ISM
absorption properties of the host galaxies \citep[e.g.,][]{fynbo05,prochaska09,sheffer09}. 
They are among the best studied events and they represent two extremes in the integrated
total ISM column density alone the afterglow line of sight.
GRB\,021004 is one of a few GRBs arising in a Lyman limit
absorber with $N_{\rm H I} = 10^{19.5\pm0.5}$ cm$^{-2}$, and the host of GRB\,080607 is a damped Ly$\alpha$ 
absorber with an unprecedentedly high gas density of $N_{\rm H I} = 10^{22.70\pm0.15}$ cm$^{-2}$.
GRB\,080607 is a dark burst with highly extinguished afterglow ($A_V\sim3.2$, \citealp{prochaska09,perley11}),
and the extinction suggests that the host galaxy may have detectable submillimeter dust emission.
The host galaxies of the two GRBs are found to have optical SFRs of 10--40 $M_\sun$~yr$^{-1}$ \citep{jakobsson05,castro10,chen11a}, 
consistent with normal star forming galaxies at $z>2$. Here we use the ALMA results to estimate the
infrared luminosities of the two host galaxies, and to examine whether there exist highly obscured 
star-forming regions that are not revealed by optical observations.  
We describe our ALMA observations and data reduction in Section~\ref{obs}, 
and the results and estimate the infrared luminosities and the SFRs of the GRB host galaxies
in Section~\ref{result}.  The implication of our observations is discussed in Section~\ref{discussion}.  
We adopt cosmological parameters of $H_0=71$ km s$^{-1}$ Mpc$^{-1}$, 
$\Omega_M=0.27$, and $\Omega_\Lambda=0.73$, 
and we convert the previous results to this set of cosmology.

\section{Observation and Data Reduction}\label{obs}
Observations of the continuum emission at 345 GHz from the host galaxies of 
GRB\,021004 and GRB\,080607 ware obtained using the ALMA 12-m array.  
Four spectral windows were tuned to center at
338, 340, 350, and 352 GHz, each with a 2 GHz bandwidth.  Bandpass calibrators and flux calibrators
were observed prior to the observations of the science targets.  Bright quasars near the GRB fields were
observed every $\sim11$ minutes for phase and amplitude calibrations.  For each science target,
a total of 0.7--0.8 hr of on-target integration was collected.
Table~\ref{tab1} summarizes the basic observing parameters and the various calibrators.

We received the data from 
the Joint ALMA Observatory (JAO)
a few weeks after the observations.  The delivered data were already bandpass, flux, and
gain (phase and amplitude) calibrated by JAO, and reference images were also provided.
All the above calibration and imaging were carried out using Common Astronomy Software 
Applications \citep[CASA,][]{mcmullin07}.  We further inspected the JAO calibration in CASA, 
and Fourier-transformed the complex visibility to make our own images.  
To obtain the highest S/N, we gave all visibility data equal weights regardless of their 
density distribution in the $uv$ plane (i.e., ``natural weighting'').
The resulting synthesized beams and sensitivities are summarized in
Table~\ref{tab1}.  We do not detect the host galaxy of GRB\,021004, and thus the imaging 
remains in the ``dirty'' stage.  We detect the host galaxy of GRB\,080607, and therefore
``CLEANed'' the sidelobes of the detected object in CASA.

\begin{deluxetable}{lcc}
\tablecaption{Observing Log\label{tab1}}
\tablehead{\colhead{Target}  & \colhead{GRB\,021004}  & \colhead{GRB\,080607}}
\startdata
Observing Date		& Oct 22, 2011			& Nov 16, 2011 \\
					& Nov 5, 2011			& Jan 12, 2012 \\
Number of Antennas\tablenotemark{a} 	& 17		& 20 \\
On-Target Integration	& 42.6 min			& 48.2 min \\
Bandpass Calibrator 	& 3c454.3 			& 3c273 \\
Flux Calibrator 			& Callisto 				& Titan \\
Gain Calibrators 		& B0007+016			& J1239+075\\
					& J0010+109			& B1236+077\\
Sensitivity (1~$\sigma$) 	& 0.113 mJy			& 0.098 mJy\\
Synthesized Beam 		& $1\farcs55 \times 1\farcs25$, $-20\arcdeg$ & $1\farcs56 \times 0\farcs87$, $3\arcdeg$\\
(major $\times$ minor, PA)
\enddata
\tablenotetext{a}{Not all antennas are used all the time.}
\end{deluxetable}

\begin{figure}[h!]
\plotone{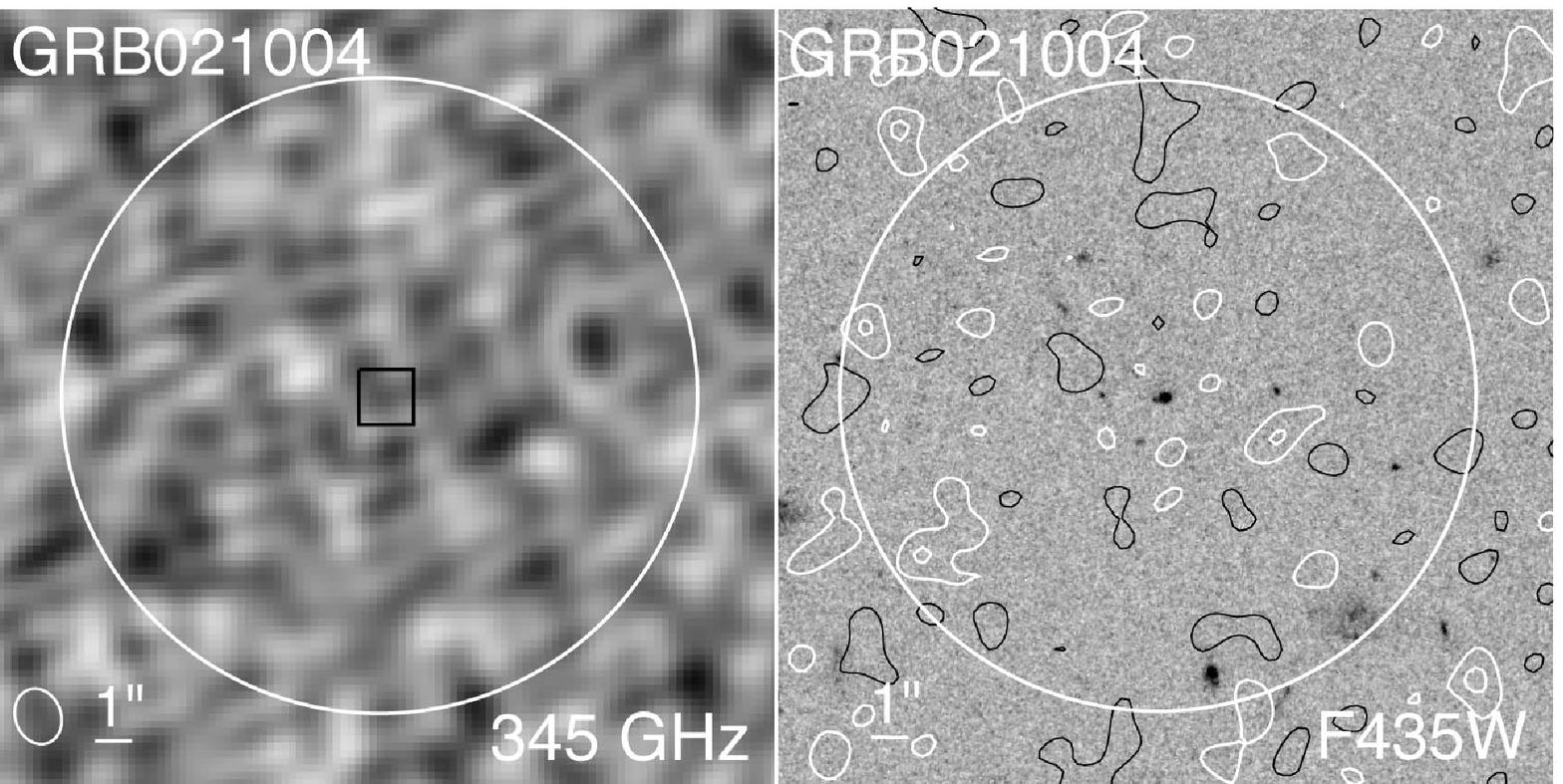}
\plotone{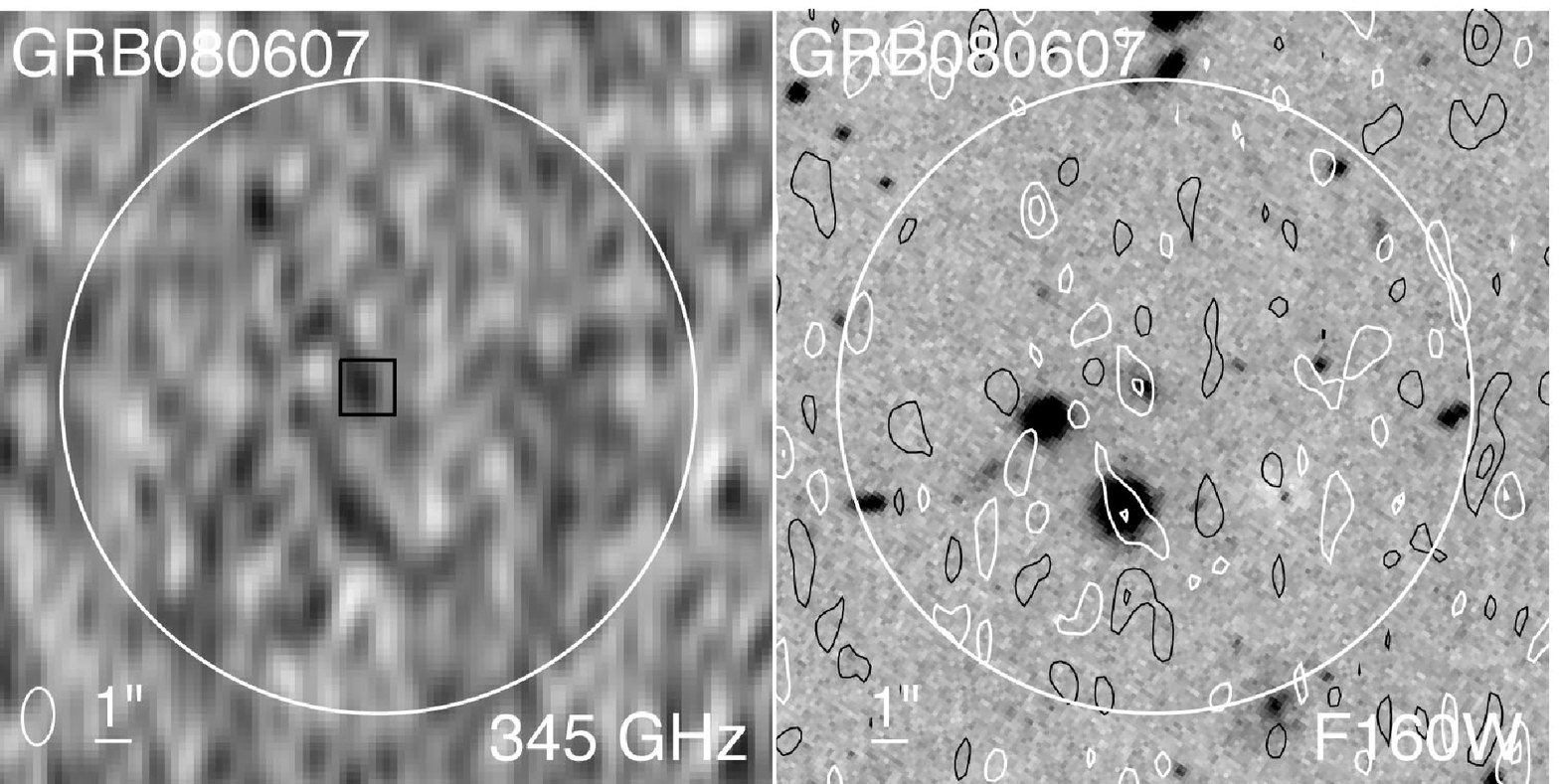}
\caption{ALMA (left) and \emph{HST} (right) images of the GRB host galaxies. In all panels, the images have inverted brightness 
scales.  The large circles indicate the 
FWHM of the ALMA primary beam, which is $17\farcs4$. In the ALMA 345 GHz panels, black boxes of 
$1\farcs5$ sizes indicate the optical positions of the host galaxies, and ellipses in the lower left indicate
the synthesized beams. In the \emph{HST} panels (ACS F435W for GRB\,021004 and WFC3 F160W for
GRB\,080607), contours of 345 GHz flux are overlaid.  White contours are +1.5 and +3.0~$\sigma$ and black 
contours are -1.5 and -3.0~$\sigma$, where 1~$\sigma$ is 0.113 and 0.098 mJy beam$^{-1}$ for GRB\,021004
and GRB\,080607, respectively.
\label{fig1}} 
\end{figure}

\section{Results}\label{result}
\subsection{GRB\,021004}\label{sec_GRB021004}
We do not detect the host galaxy of GRB\,021004 (Figure~\ref{fig1}).  
The 345 GHz point-source flux measured at the location of the host galaxy is $0.17\pm0.11$ mJy.
\citet{tanvir04} measured an 850 $\mu$m flux of $0.77 \pm 1.25$ mJy using SCUBA on JCMT.  
\citet{smith05} improved the previous SCUBA result slightly to $-1.4\pm1.0$ mJy.
Our measurement is $10\times$ deeper than these previous observations but 
the host galaxy remains undetected.

\begin{figure}
\epsscale{1.1}
\plotone{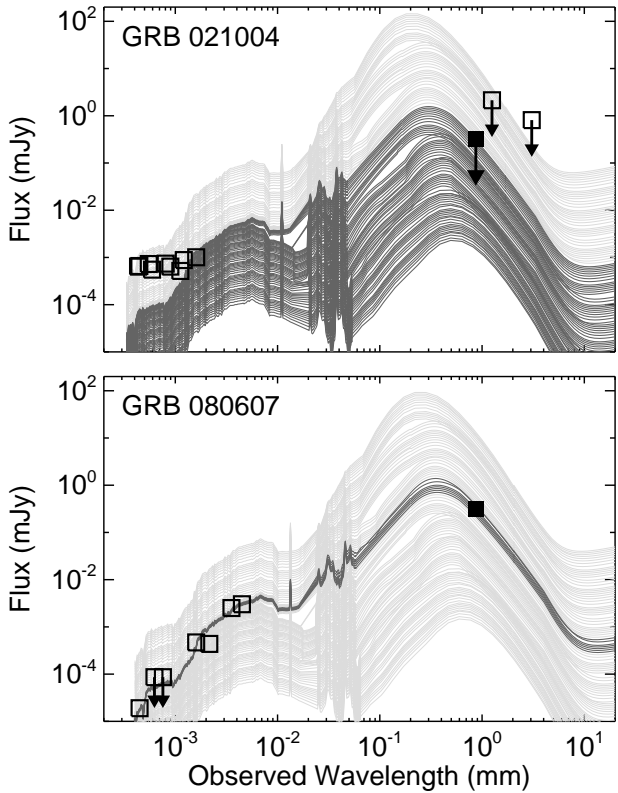}
\caption{ALMA constraints (solid squares) on the infrared SEDs of the GRB hosts. The curves are the series of 
105 redshifted CE01 SED templates, which have infrared luminosities from $2\times10^8$ to $4\times10^{13}~L_\sun$.
The thick curves are the CE01 templates that satisfy our ALMA measurements.
We also show the multiwavelength photometry from the literature (open squares; 
GRB\,021004: \citealp{fynbo05,deugartepostigo05}; GRB\,080607: \citealp{chen10,chen11a}) but we do not use them 
in the SED fitting.  Error bars are all smaller than the symbols and 
all upper limits are 3 $\sigma$.
\label{fig2}} 
\end{figure}

Given the redshift of $z=2.330$ and adopting the 3-$\sigma$ upper limit of 0.33 mJy, we can estimate 
the upper limits of its rest-frame infrared luminosity and SFR.  
To do so with single-band photometry, we need to assume a dust temperature, and this can be 
done using the infrared spectral energy distribution (SED) library of \citet[hereafer CE01]{chary01}.  
The SEDs in CE01 are luminosity dependent (based on a locally calibrated luminosity---dust temperature relation) 
and do not allow for scaling of the SEDs.  The library contains a broad range of infrared luminosity, from
$2\times10^8$ to $4\times10^{13}~L_\sun$, and each template has its unique dust temperature.
We thus redshift the CE01 SEDs to $z=2.330$ and look for those with observed 345 GHz fluxes  
below our upper limit.  Of the 105 templates provided by CE01, 65 have
345 GHz fluxes lower than 0.33 mJy (Figure~\ref{fig2}), with corresponding infrared luminosities between $2.6\times10^8$ and 
$3.1\times10^{11}$ $L_\sun$.  
An ultraluminous infrared galaxy (ULIRG, $L_{\rm IR}>10^{12}~L_\sun$)
is clearly ruled out for the host of GRB\,021004, which can be at most a 
modest infrared luminous galaxy of $L_{\rm IR}\sim3\times10^{11}$ $L_\sun$.
Combining available optical photometric measurements (open squares in Figure~\ref{fig2}) with the ALMA upper limit 
further reveals a blue SED that is inconsistent with any of the CE01 templates, suggesting that the host of
GRB\,021004 contains primarily young stars with little dust \citep[e.g.,][]{chen09} and that the infrared
luminosity may be substantially lower than the observed limit.
If we adopt the SFR conversion of star-forming galaxies, 
SFR ($M_\sun$ yr$^{-1}$) $= 1.7\times10^{-10}$ $L_{\rm IR}/L_\sun$ \citep{kennicutt98},
then the 3-$\sigma$ upper limit of the SFR of the host galaxy is 53 $M_\sun$ yr$^{-1}$.
These results are summarized in Table~\ref{tab2}.  

\citet{castro10} obtained an optical spectrum of the early-time afterglow of GRB\,021004. 
The authors estimated an unobscured SFR of $\sim40$ $M_\sun$ yr$^{-1}$ based on
the observed H$\alpha$ flux.  
\citet{jakobsson05} obtained a Ly$\alpha$ spectrum of the host galaxy and estimated an SFR
of $\sim11$ $M_\sun$ yr$^{-1}$, but this is  uncertain because of the complex radiative
transfer of Ly$\alpha$. Both results are within our 3-$\sigma$ upper limit. 
The ALMA imaging observation confirms the low SFR of the host galaxy, and also shows that more
sensitive, multi-wavelength ALMA submillimeter imaging is needed to constrain the 
infrared luminosity and dust SED of this object.

\subsection{GRB\,080607}\label{sec_GRB080607}
A significant 345 GHz flux is detected at the location of GRB\,080607.
Figure~\ref{fig1} shows our ALMA image of the field around GRB\,080607, 
and its 345 GHz flux contours overlaid on an \emph{HST} WFC3 F160W image.
The rms noise is measured to be 0.094 mJy beam$^{-1}$ within the primary beam.
There exists a peak of 345 GHz emission at the location of the host galaxy, with a peak
flux of 0.27 mJy beam$^{-1}$.  We measure a flux of 0.31 mJy by fitting the emission with a 
point-source model.  On the other hand, the contours in Figure~\ref{fig1} suggest that the
emission is elongated, and the elongation is similar to that observed in the \emph{HST} image.
We therefore also fit the emission with an extended 2-D Gaussian, and obtain a slightly higher integrated flux of 0.32 mJy.  
However, the fitted Gaussian is still consistent with a point source, which is not surprising given the low S/N.
In the subsequent 
analyses, we adopt the more conservative measurement of 0.31 mJy with a statistical significance 
of 3.3 $\sigma$.  Approximately $3\farcs5$ south of the GRB host galaxy, a marginal ($\sim3\sigma$) 
submillimeter emission is also detected at an optically bright \ion{Mg}{2}
absorber at $z=1.3399$ (H.-W. Chen et al.\ 2012, in preparation).  
We do not give further consideration to this object in this paper, but we note that this 
\ion{Mg}{2} absorber could also be a faint submillimeter source.

The key question here is whether we can consider the $\sim3.3\sigma$ emission as a detection
of the GRB host galaxy.  First, the fitted 345 GHz peak in both the point-source and the Gaussian cases 
has an offset of $0\farcs08$ from the centroid of the optical emission.  This offset is negligible given 
the S/N and the synthesized beam size of $1\farcs56\times0\farcs87$.  Thus the confidence of the 
detection is enhanced by its coincident position with the GRB host galaxy.

\begin{figure}
\epsscale{1.1}
\plotone{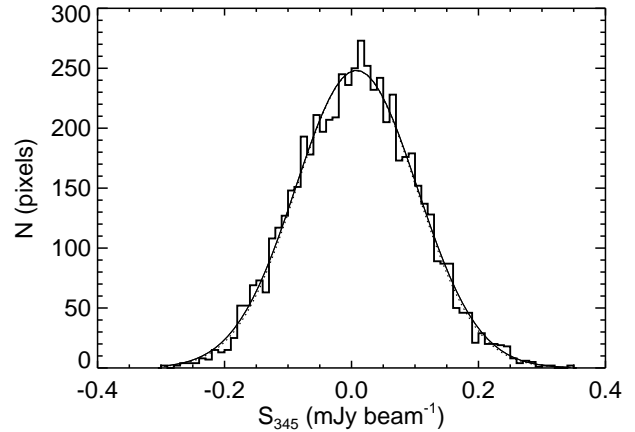}
\caption{Flux distribution of the GRB\,080607 ALMA image within the primary beam.
The histogram is fitted with a Gaussian with $\sigma=0.096$ mJy beam$^{-1}$ (solid curve), consistent with 
our measured noise of 0.094 mJy beam$^{-1}$ (dotted curve). 
\label{fig3}} 
\end{figure}

Second, we consider the probability for this peak to be spurious.
Figure~\ref{fig3} shows the histogram of pixel brightness in the primary beam. The distribution can be 
well fitted with a Gaussian with $\sigma=0.096$ mJy beam$^{-1}$ (solid curve), consistent with 
our measured noise of 0.094 mJy beam$^{-1}$ (dotted curve). 
Following the Gaussian distribution function, the probability of finding a $>3.3\sigma$ noise 
spike is $5\times10^{-4}$.  However, there are hints of a non-Gaussian noise.  
The histogram in Figure~\ref{fig3} suggests an excess of positive pixels (at $>0.2$ mJy). 
In the image (Figure~\ref{fig1}), additional to the GRB host galaxy, there is a 
second $>3.3\sigma$ spike to the north-east of
the GRB host, which has no known optical counterpart.  Within the ALMA primary beam 
(FWHM = $17\farcs4$), there are approximately 220 independent resolution elements.  
This additional $>3.3\sigma$ spike suggests a probability of $1/220 = 5\times10^{-3}$, 
which is $10\times$ higher than the Gaussian probability.  This is an upper limit, since we cannot 
rule out this spike as a real submillimeter source. Thus the probability of finding a 
$>3.3\sigma$ spike at the location of our target is between $5\times10^{-4}$ (assuming a Gaussian noise) 
and $5\times10^{-3}$ (assuming that the second 3.3 $\sigma$ spike is due to noise). 
Both these values are sufficiently small.   Therefore, the fact that the observed emission coincides 
with the position of the GRB host substantially increases the confidence level of the detection of the host. 
In this paper, we consider this a detection, but we also point out that it will be worthwhile to confirm 
this  with ALMA in future larger GRB host galaxy surveys.

With the above measured flux and the redshift of $z=3.036$, we then estimate the infrared luminosity 
of the host galaxy of GRB\,080607, using the CE01 library.
We redshift the CE01 templates to $z=3.036$ and find seven of the 105 templates fall in
the observed range of $0.313\pm0.094$ mJy (Figure~\ref{fig2}), with
$L_{\rm IR} = 2.4\times10^{11}$ to $4.5\times10^{11}$ $L_\sun$.
Including available optical and near-infrared photometric measurements, the bottom panel of
Figure~\ref{fig2} further shows that the SED is well represented by known dusty templates of local 
galaxies across the full spectral range.  The agreement strongly supports the 
conclusion that the host galaxy of 
GRB\,080607 is similar to a luminous infrared galaxy (LIRG, 
$L_{\rm}>10^{11}~L_\sun$) in the local universe. The inferred SFR is between 41 and 77 $M_\sun$ yr$^{-1}$.  
The above results are summarized in Table~\ref{tab2}.

\citet{chen11a} presented SED fitting for the host of GRB\,080607 at 
$\lambda_{\rm rest} \sim 0.4$--4 $\mu$m.  Adopting a Milky-Way type dust extinction law, they 
found $A_V=1.24$ and an extinction corrected
SFR of $\sim16$--24 $M_\sun$ yr$^{-1}$, roughly $3\times$ lower than the submillimeter SFR.  
Given known uncertainties in both the optical extinction
correction and the infrared SED of galaxies and  possible variation in the distribution of dust content, 
the factor of three difference between the optical and submillimeter SFRs only suggests 
a modest amount of dust enshrouded star formation.

\begin{deluxetable}{lcccc}
\tablecaption{Properties of the GRB Host Galaxies\label{tab2}}
\tablehead{\colhead{Target}  & \colhead{Redshift}  & \colhead{$S_{\rm345}$ (mJy)}  & \colhead{$L_{\rm IR}$ ($10^{11}$ $L_\sun$)} & \colhead{SFR ($M_\sun$ yr$^{-1}$)}}
\startdata
GRB\,021004\tablenotemark{a}	& 2.330	& $<0.33$			& $<3.1$		& $<53$ \\
GRB\,080607 					& 3.036	& $0.31\pm0.09$	& 2.4--4.5		& 41--77
\enddata
\tablenotetext{a}{The upper limits for GRB\,021004 are all 3 $\sigma$.}
\end{deluxetable}

\section{Discussion}\label{discussion}
With the pilot ALMA imaging program of two GRB host galaxies at $z>2$, we attempt to 
constrain the far-infrared properties of GRB host galaxies. In our sample, GRB\,021004 has a very bright
afterglow, but the host galaxy does not appear to show unusual dust content.
On the other hand, GRB\,080607 is a dark GRB with large extinction along the line of sight.
Their host galaxies have measured 345 GHz fluxes of $0.17\pm0.11$ and $0.31\pm0.09$ mJy, respectively. 
Statistically, we cannot rule out the possibility that the two host galaxies have 
comparable submillimeter fluxes.  
Despite that we had already pushed the sensitivities to roughly an order of magnitude 
deeper than previous measurements, deeper ALMA observations (and a larger sample) are 
clearly needed to tell the difference between the host galaxies of typical and dark GRBs.  

On the other hand, the ALMA sensitivity limit is deep enough to probe beyond the ULIRG regime.
\citet{chen10}  suggest that the high-redshift 
infrared luminous galaxy population contributes to the GRB host galaxy population.  The ALMA sensitivity 
thus allows us to examine whether the two host galaxies are similar to typical dusty galaxies selected 
by submillimeter telescopes (submillimeter galaxies, hereafter SMGs).
First, it is established that bright 345 GHz selected SMGs primarily reside in the redshift range 
of $z=1.5$--3.5 \citep{barger00,chapman03,chapman05,wardlow11}. 
The host galaxies of GRB\,021004 and GRB\,080607 have redshifts in the range
of these bright SMGs.  
The integrated source counts indicate that bright SMGs of $S_{345}>2$ mJy contribute to $\sim30\%$
of the extragalactic background light in this wavelength range \citep[e.g.,][]{coppin06}.
Faint-end counts derived from lensing cluster surveys indicate that approximately 50\% of the background arises from
fainter sources in the flux range of $S_{345}=0.5$--2 mJy \citep{cowie02,knudsen08,chen11b}.
Our ALMA detection and tight upper limit on the host galaxies of GRB\,021004 and
GRB\,080607 thus put them at the still fainter end of the 345 GHz population.

We further compare the two GRB host galaxies with normal star forming galaxies at high redshift.
There exists a correlation between SFR and stellar mass of star forming galaxies 
at high redshift \citep[e.g.,][]{daddi07,pannella09,karim11,rodighiero11}.
This correlation is often referred to as the star formation ``main sequence'' of galaxies.  Galaxies at the
main sequence are suggested to be disks that undergo quasi-steady star formation, and outliers
are suggested to be starbursts with star formation boosted by gas-rich mergers \citep{daddi10,genzel10}.
For $z\sim2$, \citet{daddi07} found SFR = 200 $M_{11}^{0.9}$ ($M_\sun$ yr$^{-1}$) for main-sequence
galaxies, where $M_{11}$ is stellar mass in units of $10^{11} M_\sun$. The host galaxy of GRB\,021004 has
estimated stellar masses of $1.6\times10^{10} M_\sun$ \citep{savaglio09} and  $2.6\times10^{9} M_\sun$ \citep{chen09}.
With our ALMA SFR upper limit ($3\sigma$) of 53 $M_\sun$ yr$^{-1}$, its SFR/$M_{11}^{0.9}$ has values of 
$<280$ or $<1400$, depending on the adopted stellar mass.  The former is consistent with main-sequence
galaxies, while the latter is close to a starburst.  However, both values are 3-$\sigma$ upper limits.
The host galaxy of GRB\,080607 has a better constrained stellar mass of  1--3 $\times10^{10} M_\sun$ \citep{chen11a}.  
If we adopt our ALMA SFR of 41--77 $M_\sun$ yr$^{-1}$, then it has SFR/$M_{11}^{0.9}=100$--600.
This exercise shows that both GRB host galaxies are consistent with being main-sequence star-forming galaxies.

Finally, the afterglow spectrum of GRB\,080607 shows a fairly large dust extinction of $A_V=3.2$, and
unprecedentedly high gas densities of $N_{\rm H I} = 10^{22.70\pm0.15}$ cm$^{-2}$ 
and $N_{\rm CO} = 10^{16.5\pm0.3}$  cm$^{-2}$, with a warm CO excitation
temperature of $T_{\rm ex}^{\rm CO} > 100$ K \citep{prochaska09}.
However, Prochaska et al. also suggest that the intervening molecular cloud is not the birth place of
the GRB. The extinction in the afterglow is significantly larger than that for the host galaxy 
($A_V=1.2$, \citealp{chen11a}). All the above, together with our ALMA result of a relatively normal SFR, 
indicates that GRB\,080607 is not in a rare dusty galaxy, but the sightline happens to pass through 
a molecular cloud in its host galaxy.

\acknowledgments
We thank the referee for the useful comments.
This paper makes use of the  ALMA data: ADS/JAO.ALMA\#2011.0.00101.S. 
ALMA is a partnership of ESO (representing its member states), NSF (USA) and NINS 
(Japan), together with NRC (Canada) and NSC and ASIAA (Taiwan), in cooperation with 
the Republic of Chile. The JAO is operated by ESO, AUI/NRAO and NAOJ.
This work is supported by the NSC grant 99-2112-M-001-012-MY3 (W.H.W.).

\end{document}